\documentclass[sort&compress]{elsarticle}

\usepackage{amsmath,amssymb,amsfonts}
\usepackage{graphicx}
\usepackage{subcaption}
 
\usepackage{numcompress}

\newcommand{\Fig}[1]{Fig.~(\ref{#1})}
\newcommand{\Eq}[1]{Eq.~(\ref{#1})}
\newcommand{\eq}[1]{Eq.~(\ref{#1})}

\newcommand{\Order}[1]{O(#1)}

\begin{document}

\title{Mechanically-driven spreading of bacterial populations}

\author[up]{Waipot Ngamsaad\corref{cor1}}
\ead{waipot.ng@up.ac.th}
\address[up]{Division of Physics, School of Science, University of Phayao, Phayao 56000, Thailand}
\cortext[cor1]{Corresponding author}

\author[cmu]{Suthep Suantai}
\ead{suthep.s@cmu.ac.th}
\address[cmu]{Department of Mathematics, Faculty of Science, Chiang Mai University, Chiang Mai 50200, Thailand}

\date{\today}

\begin{abstract}
The effect of mechanical interactions between cells in the spreading of bacterial populations was investigated in one-dimensional space. A continuum-mechanics approach, comprising cell migration, proliferation, and exclusion processes, was employed to elucidate the dynamics. The consequent nonlinear reaction-diffusion-like equation describes the constitution dynamics of a bacterial population. In this model, bacterial cells were treated as rod-like particles that interact with each other through hard-core repulsion, which introduces the exclusion effect that causes bacterial populations to migrate quickly and at high density. The propagation of bacterial density as a traveling wave front over extended times was also analysed. The analytical and numerical solutions revealed that the front speed was enhanced by the exclusion process, which depended upon the cell-packing fraction. Finally, we qualitatively compared our theoretical results with experimental evidence.
\end{abstract}


\maketitle

\section{Introduction\label{sec:Intro}}
In recent decades, much attention has been paid to the collective behavior of bacterial populations. This system is used as the prototype for understanding multicellular assemblies, such as tissue and biofilm \cite{Shapiro1998}. The insight into the underlying mechanism of dynamics is important to biological and medical science. 

To cope with unfavorable environmental conditions, bacterial colonies generate varieties of pattern formations \cite{Kawasaki1997, Ben-Jacob2000}. The spatiotemporal pattern formation in bacterial colonies results from cell migration and proliferation. These dynamics at a continuum level can be described by reaction-diffusion processes \cite{Kawasaki1997, Golding1998, Ben-Jacob2000}. The simplified model \cite{Kawasaki1997} relied on a density-dependent (or degenerate) reaction-diffusion equation \cite{Gurney1975, Gurtin1977, Newman1980, Newman1983, Murray1989}, which was an extension of the classical Fisher-KPP equation \cite{Fisher1937, Tikhomirov1991}. These well-known solutions \cite{Newman1980, Newman1983} revealed that bacterial density evolves as a sharp traveling wave with constant front speed \cite{Kawasaki1997}. In our previous work, we found an explicit space-time solution for the generalized Fisher-KPP equation in one-dimensional space \cite{Ngamsaad2012}. This solution evolves from a specific initial condition to a self-similar object that converges to the usual traveling wave on an extended time scale. Although capable of explaining these dynamics, the conventional model omitted the size of the bacterial cell. In real systems, most bacterial cells are rod shaped and grow in dense environments. Accordingly, the mechanical interactions between cells could play crucial roles in the spreading of bacterial colonies. 

Recent experimental and theoretical studies showed that mechanical interactions between cells have important roles in the collective behavior of bacterial colonies \cite{Cho2007, Volfson2008, Mather2010, Boyer2011, Su2012, Grant2014}. The dependence on the elastic modulus of the front speed has theoretically been found \cite{Farrell2013}. It mentions that \emph{the migration of bacteria is caused by cell pushing rather than self-propulsion} in dense colonies \cite{Volfson2008, Su2012, Grant2014}. Therefore, we speculate that the exclusion process that prevents the overlapping of cells could play a crucial role in the spreading of bacterial colonies. This issue motivates us to extend the conventional density-dependent reaction-diffusion equation \cite{Gurney1975, Gurtin1977, Newman1980, Newman1983, Murray1989} by incorporating the cell size into the investigation of the dynamics of bacterial populations.

In this work, we considered the systems of bacterial cells growing on a thin layer of nutrient-rich fluid medium. The bacteria increased in population through cell division and interacted through hard-core repulsion (steric interactions), which resulted in exclusion effects and consequent non-overlapping of cells. Although bacteria are self-propelled particles \cite{Cates2012}, in colonies of densely packed or non-motile cells, bacterial migration was caused by cell pushing, resulting from cell growth and division, rather than self-propulsion \cite{Volfson2008, Su2012, Grant2014}. Thus, \emph{the bacteria behave similarly as passive particles or nonmotile cells} in high density environments. Apart from cells, Bruna and Chapman \cite{Bruna2012b} observed that the self-diffusion of hard, spherical Brownian particles in a dilute regime decreased as the density increased, due to the diffusion of any single particle being impeded by collisions with other particles. However, these collisions encouraged the particle to move toward low-density regions, resulting in this biased migration being faster than self-diffusion and enhancing overall collective diffusion. Guided by the work of Bruna and Chapman \cite{Bruna2012b}, we propose that bacterial cells move based on hard-core repulsion and without self-propelled motility in dense colonies.

After incorporating exclusion processes in cell (or particle) dynamics, altered diffusion coefficients in the continuum limits were found \cite{Bodnar2005, Lushnikov2008, Simpson2011, Baker2011, Bruna2012, Dyson2012, Penington2014, Almet2015}. The enhancement or slowing of diffusion depends upon cell length and the available moving distance, as shown by lattice-based analysis \cite{Penington2014}. In some models, diffusion diverges to infinity in closely packed densities \cite{Bodnar2005, Lushnikov2008, Almet2015}. Singular diffusion has also been modeled through the migration of bacterial biofilm \cite{Eberl2007, Jalbert2014} and glioblastoma tumors \cite{Harko2015}. However, the effect of diverged diffusion on the propagation speed of cell populations remains unknown.

To address this question, we employed a continuum-mechanics approach to cell proliferation \cite{Arciero2011} in order to investigate the spreading of bacterial populations in the presence of exclusion processes. Additionally, we analytically and numerically elucidated the front speed of bacterial colony expansion in terms of cell size and discussed the consistency of our theoretical results with the experimental evidence.


\section{Continuum mechanical model \label{sec:Model}}

\subsection{Constitution equations}
From a macroscopic view, bacterial populations constitute continuum fluid capable of reproducing in order to increase cell numbers. By pushing each other following cell division \cite{Volfson2008, Su2012, Grant2014}, population pressure increases as a result of collisions between cells and forces cells to move. During movement, cells encounter friction from the surrounding fluid medium and the substrate surface. For the sake of simplicity, we considered the expansion of bacterial colonies in one-dimensional space, regardless of cell orientation. Adapting from \cite{Arciero2011}, the constitution equations that describe the evolution of the cell density, $\rho(x,t)$, and collective velocity, $V(x,t)$, of the bacterial population at position $x$ and time $t$ are given by
\begin{eqnarray}
\label{eq:continue}
\frac{\partial \rho}{\partial t}  &=& - \frac{\partial \left(\rho V\right)}{\partial x} + \Gamma(\rho), \\
\label{eq:friction}
-\gamma V &=& \frac{\partial p}{\partial x} =  \frac{\partial p}{\partial \rho}\frac{\partial \rho}{\partial x},
\end{eqnarray}
where $\Gamma(\rho(x,t))$ represents the growth function, $p(\rho(x,t))$ represents the internal population pressure, and $\gamma$ represents the damping constant. \Eq{eq:continue} represents the continuity equation with the growth term. We assume that bacterial growth obeys the law of population growth as described by a logistic function: $\Gamma(\rho) = k\rho\left(1-\rho/\rho_m\right)$, where $k$ is the rate constant and $\rho_m$ is the maximum density \cite{Murray1989, Arciero2011}. \Eq{eq:friction} arises from the force balance between Stokes' law for friction and the pressure gradient, which is similar to Darcy's law describing fluid flow through a porous medium.

We model the bacterial cells as non-overlapping hard-rod particles of average length, $\sigma$, that interact through hard-core repulsion. In high-density environments, bacterial self-propulsion can be ignored, since it is dominated by collision between cells. This defines the bacterial cell as a passive particle or non-motile cell that obeys the laws of thermodynamics. For hard-rod fluid in one dimension, the exact pressure is given as
\begin{equation}\label{eq:Tonks_pressure}
p(\rho) = \frac{\rho k_B T}{1-\sigma\rho},
\end{equation}
where $k_B$ is the Boltzmann constant and $T$ represents the temperature \cite{Tonks1936, Salsburg1953, Helfand1961}. In our case where bacterial cells behave as passive particles, the temperature relates to the average translational kinetic energy of a cell, $<E_k>=(1/2)k_BT$, we assume that the temperature is constant in our system. The pressure in \eq{eq:Tonks_pressure} diverges to infinity at closely packed density: $\rho\to 1/\sigma$. Notably, in dilute density, $\rho\to 0$, \eq{eq:Tonks_pressure} recovers the pressure of an ideal gas: $p = \rho k_B T$. As shown by \cite{Takatori2014, Takatori2015, Solon2015}, the pressure for dilute active particles is similar to the ideal gas, except that the source of kinetic energy comes from the swim speed, $U_0$: $k_B T \propto U_0^2$ \cite{Takatori2014, Takatori2015}. As will be shown later, the temperature source is not important; as long as it is constant, the dynamics of our model are invariant.

\subsection{Dimensionless equations \label{sec:Dimless}}		
We define the maximum density as $\rho_m = 1/\sigma_m$, where $\sigma_m$ represents the average length occupied by one cell and $\sigma_m > \sigma > 0$. The logistic law limits the growth of bacteria, such that $0\leq\rho\leq\rho_m < 1/\sigma$. For convenience of further analysis, we introduce the following dimensionless quantities:  $0\leq u=\rho/\rho_m\leq 1$, $v=[\gamma/(k\rho_mk_BT)]^{1/2}V$, $0 < \epsilon=\sigma\rho_m=\sigma/\sigma_m < 1$, $t^\prime=\alpha t$, and $x^\prime=[(k\gamma)/(\rho_mk_BT)]^{1/2}x$. In one dimension, the packing fraction, ($\epsilon$), represents the length fraction, which is equivalent to the area and volume fractions in two and three dimensions, respectively. We then rewrite \Eq{eq:continue} and \Eq{eq:friction} by employing \eq{eq:Tonks_pressure} in dimensionless form:
\begin{eqnarray}
\label{eq:continue2}
\frac{\partial u}{\partial t}  &=& -\frac{\partial \left(u v\right)}{\partial x} + u(1-u), \\
\label{eq:friction2}
v &=& -\frac{1}{\left(1-\epsilon u\right)^2}\frac{\partial u}{\partial x} ,
\end{eqnarray}
where the prime has been dropped. From \eq{eq:friction2}, the migration of bacterial populations is biased to move down the density gradient and enhanced by the exclusion process, implied from the factor $1/(1-\epsilon u)^2$. This factor increases with the density and diverges to infinity as $\epsilon\to 1$ at $u=1$, which causes the bacterial population to migrate faster at higher density. This singularity has appeared in similar models using different approaches \cite{Bodnar2005, Eberl2007, Lushnikov2008, Harvey2013, Jalbert2014, Harko2015, Almet2015}. Fortunately, the velocity in \eq{eq:friction2} is finite, since $\partial u/\partial x \to 0$ at $u=1$. The density inside of the colony reaches a saturated value, except in proximity to the colony edge. In this regime, the density distribution is homogeneous and its gradient approaches zero. 

Substituting \eq{eq:friction2} into \eq{eq:continue2}, we obtain a nonlinear partial differential equation:
\begin{equation}\label{eq:cell_RD_dimensionless_gen}
\frac{\partial u}{\partial t} = \frac{\partial}{\partial x} \left(M(u)\frac{\partial u}{\partial x} \right) + g(u),
\end{equation}
where $M(u) = u/(1-\epsilon u)^2$ and $g(u) = u(1-u)$. \eq{eq:cell_RD_dimensionless_gen} is in the same form as the density-dependent reaction-diffusion equation, however, the migration and diffusion coefficients differ. This is unrelated to the mean-square displacement, however, $M \sim \rho\partial p/\partial \rho$. In this model, the populations migrate based on the collision between cells as opposed to a random walk. A similar coefficient represents the contribution of hard-core repulsion between cells to the migration of myxobacteria in a dense phase \cite{Harvey2013}. \eq{eq:cell_RD_dimensionless_gen} is degenerate based on $M(0)=0$, which results in the sharp interface separated between occupied and cell-free regions. In a very dilute system ($\epsilon\to 0$), \eq{eq:cell_RD_dimensionless_gen} recovers the conventional degenerate Fisher-KPP equation \cite{Newman1980, Newman1983, Murray1989}, for which an explicit solution was determined in our previous work \cite{Ngamsaad2012}.

\section{Traveling-wave solution \label{sec:Analysis}}		
We focused on behaviour of the system over extended times, during which the population density propagates as a traveling wave: $u(x,t)=\phi(z)$, where $z=x-ct$ and $c$ represent the front speed \cite{Murray1989}. Substituting the traveling-wave solution into \eq{eq:cell_RD_dimensionless_gen}, we obtain
\begin{equation}\label{eq:cell_RD_wave}
\frac{d}{dz} \left(M(\phi)\frac{d\phi}{dz} \right) + c\frac{d\phi}{dz} + g(\phi) = 0.
\end{equation}
In the degenerate model, the density must vanish at the finite position, $z^\ast (< \infty)$, that undergoes the sharp interface. We then consider the density profile that satisfies the following conditions: $\phi(-\infty) = 1$, $\phi(z) = 0$ for $z\geq z^\ast$, $\frac{d}{dz}\phi(-\infty) = 0$, and $\frac{d}{dz}\phi(z^\ast) \neq 0$. Additionally, for $\epsilon\in [0,1)$, $M(\phi(-\infty)) < \infty$ and $M(\phi(z)) = 0$ for $z\geq z^\ast$ \cite{Sanchezgarduno1995}. Multiplying \eq{eq:cell_RD_wave} by $M(\phi)d\phi/dz$ and then integrating with respect to $z$ from $-\infty$ to $z^\ast$, we obtain 
$c\int_{-\infty}^{z^\ast} M(\phi)\left(\frac{d\phi}{dz}\right)^2 dz   
+ \int_{-\infty}^{z^\ast} M(\phi)g(\phi)\frac{d\phi}{dz} dz + \left. \frac{1}{2} \left(M(\phi)\frac{d\phi}{dz} \right)^2 \right|_{-\infty}^{z^\ast}  = 0$.
Under these density profile conditions, the last term on the left-hand side is zero. Finally, we obtain the front speed:
\begin{equation}\label{eq:wave_vel3}
c = - \frac{\int_{0}^{1} M(\phi)g(\phi)d\phi}{ \int_{0}^{1} M(\phi)\left(\frac{d\phi}{dz}\right) d\phi }.
\end{equation}
To obtain the closed-form of the front speed, $c$, the solution for the density gradient, $d\phi/dz$, is required.

\subsection{Approximate solution}
Although the exact solution of \eq{eq:cell_RD_wave} remains unknown, we can find the approximate solution by employing the perturbation method \cite{Sanchezgarduno1994}. By defining $w(\phi) = d\phi/dz$, we rewrite \eq{eq:cell_RD_wave}:
\begin{equation}\label{eq:perturb1}
M(\phi)w\frac{dw}{d\phi} +  M^{\prime}(\phi)w^2 + cw + g(\phi) = 0,
\end{equation}
where $M^{\prime}(\phi) = dM(\phi)/d\phi$. The migration coefficient can be written in the expansion form:
$M(\phi) \approx \phi\left(1 + 2\phi\epsilon + 3\phi^2\epsilon^2 + \cdots\right)$.
We then look for the solution of \eq{eq:perturb1} in the power series of $\epsilon$:
\begin{eqnarray}
\label{eq:v_pert}
w(\phi) &=& w_0(\phi) + w_1(\phi)\epsilon + w_2(\phi)\epsilon^2 + \cdots, \\
\label{eq:c_pert}
c &=& c_0 + c_1\epsilon + c_2\epsilon^2 + \cdots,
\end{eqnarray}
where $w_i(\phi)$ and $c_i$, that $i\in \lbrace0,1,2,\ldots,\infty\rbrace$ are coefficients to be determined. Substituting \eq{eq:v_pert} and \eq{eq:c_pert} into \eq{eq:perturb1}, we obtain the equation for each order as follows: at $\epsilon^0$,
\begin{equation}\label{eq:order0}
\phi w_0 \frac{dw_0}{d\phi} + w_0^2 + c_0w_0 + \phi(1-\phi) = 0,
\end{equation}
and, at $\epsilon^1$, 
\begin{eqnarray}\label{eq:order1}
\lefteqn{
\phi w_0 \frac{dw_1}{d\phi} + \left(\phi\frac{dw_0}{d\phi} + 2w_0 + c_0\right)w_1 
}\nonumber\\&&
+ 2\phi^2w_0\frac{dw_0}{d\phi} + 4\phi w_0^2 + c_1w_0 = 0.
\end{eqnarray}
\eq{eq:order0} has the known solutions: $w_0=(1/\sqrt{2})(\phi-1)$ and $c_0=1/\sqrt{2}$ \cite{Newman1980, Newman1983, Murray1989, Sanchezgarduno1994}. Substituting these solutions into \eq{eq:order1}, we obtain a linear first-order ordinary differential equation:
\begin{eqnarray}\label{eq:ode_order1}
\lefteqn{
\phi (\phi-1) \frac{dw_1}{d\phi} + \left(3\phi - 1\right)w_1 
}\nonumber\\&&
+ 3\sqrt{2}\phi^3 -5\sqrt{2}\phi^2 +(2\sqrt{2}+c_1)\phi - c_1 = 0 .
\end{eqnarray}
After finding the integrating factor \cite{Arfken1985}, we obtain its solution:
\begin{eqnarray}\label{eq:sol_order1}
\lefteqn{
w_1(\phi) = \frac{1}{\left(\phi - 1\right)^2 } \left[\frac{C}{\phi} - \frac{3 \sqrt{2}}{5} \phi^{4} + 2 \sqrt{2} \phi^{3} \right. 
}\nonumber\\&&
\left. - \left(\frac{c_{1}}{3} + \frac{7 \sqrt{2}}{3}\right)\phi^{2}  + \left(c_{1} + \sqrt{2}\right)\phi  - c_{1}  \right],
\end{eqnarray}
where $C$ is the integral constant. To prevent the singularity at $\phi = 0$ and $\phi = 1$, we require that $C=0$ and $- \frac{3 \sqrt{2}}{5} + 2 \sqrt{2}- \left(\frac{c_{1}}{3} + \frac{7 \sqrt{2}}{3}\right)+ \left(c_{1} + \sqrt{2}\right)  - c_{1}  = 0$. Thus, we obtain
\begin{equation}\label{eq:c1}
c_1 = \frac{2}{5\sqrt{2}}.
\end{equation}
Substituting \eq{eq:c1} into \eq{eq:sol_order1}, we obtain
\begin{equation}\label{eq:v1}
w_1(\phi) = -\frac{2}{5\sqrt{2}}(\phi - 1)(3\phi - 1).
\end{equation}
Finally, gathering all terms, we obtain the approximate solutions with the correction of $\Order{\epsilon^2}$
\begin{eqnarray}
\label{eq:sol_e2}
w  = \frac{d\phi}{dz} &=& \frac{6\left(\phi - 1\right)}{5\sqrt{2}}  \left(\frac{5+2\epsilon}{6}-\epsilon\phi\right) + \Order{\epsilon^2}, \\
\label{eq:vel_e2}
c &=& \frac{1}{\sqrt{2}}\left(1 + \frac{2}{5}\epsilon\right) + \Order{\epsilon^2}.
\end{eqnarray}
The density gradient approaches zero when the density reaches the maximum value, $\phi\to 1$, as expected. By using $w(\phi)=d\phi/dz$, we can calculate the approximate density profile:
\begin{equation}\label{eq:den_pro}
\phi(z) = \left\{ 
\begin{array}{lc} 
\frac{1-\exp{\left[b(z-z_0)\right]}}{1-a\exp{\left[b(z-z_0)\right]}}, & z \leq z_0 \\ 
0, & z > z_0 ,
\end{array} \right.
\end{equation}
where $a=\frac{6\epsilon}{5+2\epsilon}$, $b=\frac{5-4\epsilon}{5\sqrt{2}}$, and $z_0$ represents the initial front position where $\phi(z_0) = 0$.

\subsection{Front speed}
The front speed is the collective velocity at the edge of the colony, $c=v(\phi(z^\ast))=v(0)$. Based on the correction of $\Order{\epsilon^2}$ from \eq{eq:vel_e2}, the front speed increases linearly with packing fraction ($\epsilon$). However, substituting \eq{eq:sol_e2} into \eq{eq:wave_vel3} and after integration, we can obtain a more precise front speed:
\begin{equation}\label{eq:wave_vel_approx}
c(\epsilon) = \frac{5}{\sqrt{2}\epsilon} \frac{\left(4\epsilon - 6\right)\ln{\left(1-\epsilon\right)} + \epsilon^2 - 6\epsilon }{\left(2\epsilon^2 - 11\epsilon + 8\right)\ln{\left(1-\epsilon\right)} - 7\epsilon^2 + 8\epsilon }.
\end{equation}
The front speed depends upon the packing fraction of a cell. Therefore, the front speed recovers the usual value, that $c_0=1/\sqrt{2}\approx 0.7071$, in a very dilute regime, as $\epsilon\to 0$ \cite{Newman1980, Newman1983, Murray1989, Sanchezgarduno1994}. In a closely packed regime, as $\epsilon\to 1$, the front speed approaches a finite value, that $c(1) = 10/\sqrt{2} \approx 7.071$, and increases by a factor of 10 from the dilute regime.

\section{Numerical results and discussion\label{sec:Results}}
As the correction of our approximate solutions is limited to $\Order{\epsilon^2}$, it is counterintuitive, given that the model is designed for capturing dynamics at high density. To obtain the actual results at high density, we solved \eq{eq:cell_RD_dimensionless_gen} directly and subjected the solution to a zero-flux boundary condition using the numerical method. In \eq{eq:cell_RD_dimensionless_gen}, the migration coefficient increases with density, which is inefficient when solving with an explicit finite-difference scheme \cite{NumericalRecipes}. Unfortunately, solving with the standard implicit-numerical scheme is also difficult because of the factor $1/\left(1-\epsilon u\right)^2$. We found that the simplest algorithm that overcomes these obstructions is the nonstandard fully implicit finite-difference method \cite{Eberl2007}. This algorithm has proven stable enough to explore the dynamics at high-packing fractions. The detailed algorithm is described in the Appendix.

Although our model is not expected to be accurate for dilute systems, since it has neglected bacterial self-propulsion, we focused on bacterial population dynamics by varying the values of the packing fraction, $\epsilon\in[0,1)$. For our computation, we chose the spacing step and the time step, such that $\delta x = 0.05$ and $\delta t = 0.01$, respectively. The computations were performed on 3000 grids for $\epsilon \in [0,0.5]$ and on 5000 grids for $\epsilon \in (0.5,0.99]$, with 8000 iterations. For $\epsilon=0.999999$, the computation was performed on 120,000 grids with 150,000 iterations. The initial density profile, $u_0(x)$, was set to a step function:
\begin{equation}\label{eq:step_fn}
u_0(x) = \left\{ 
\begin{array}{lc} 
1, & x < r_0 \\ 
0, & x \geq r_0 ,
\end{array} \right.
\end{equation}
where $r_0$ represents the initial front position. To ensure that it was far enough from the boundary at origin, we set $r_0 = 50$. 

\begin{figure}[ht]
\begin{subfigure}{0.5\textwidth}
\centerline{\includegraphics[width=\columnwidth]{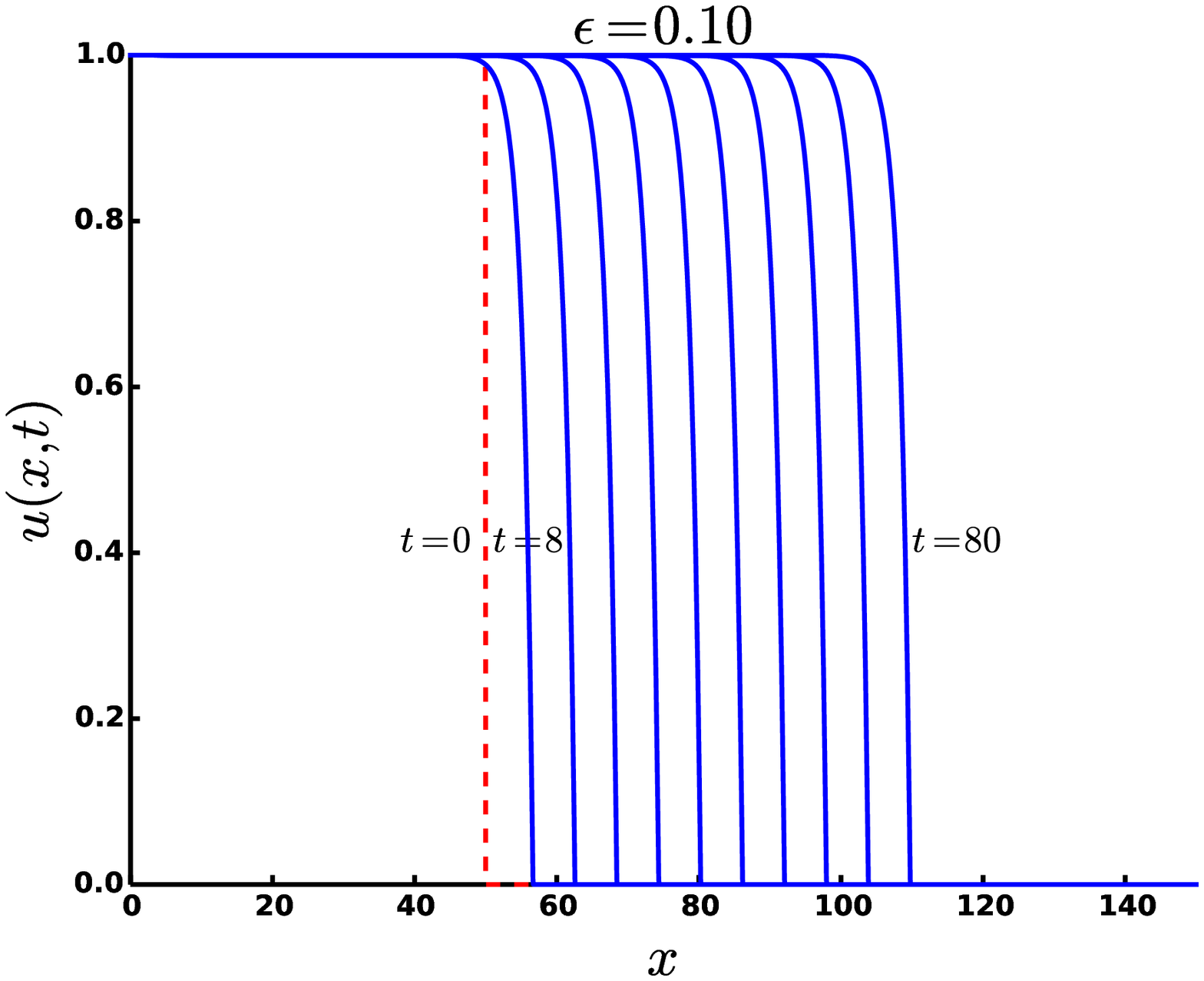}}
\end{subfigure}
\begin{subfigure}{0.5\textwidth}
\centerline{\includegraphics[width=\columnwidth]{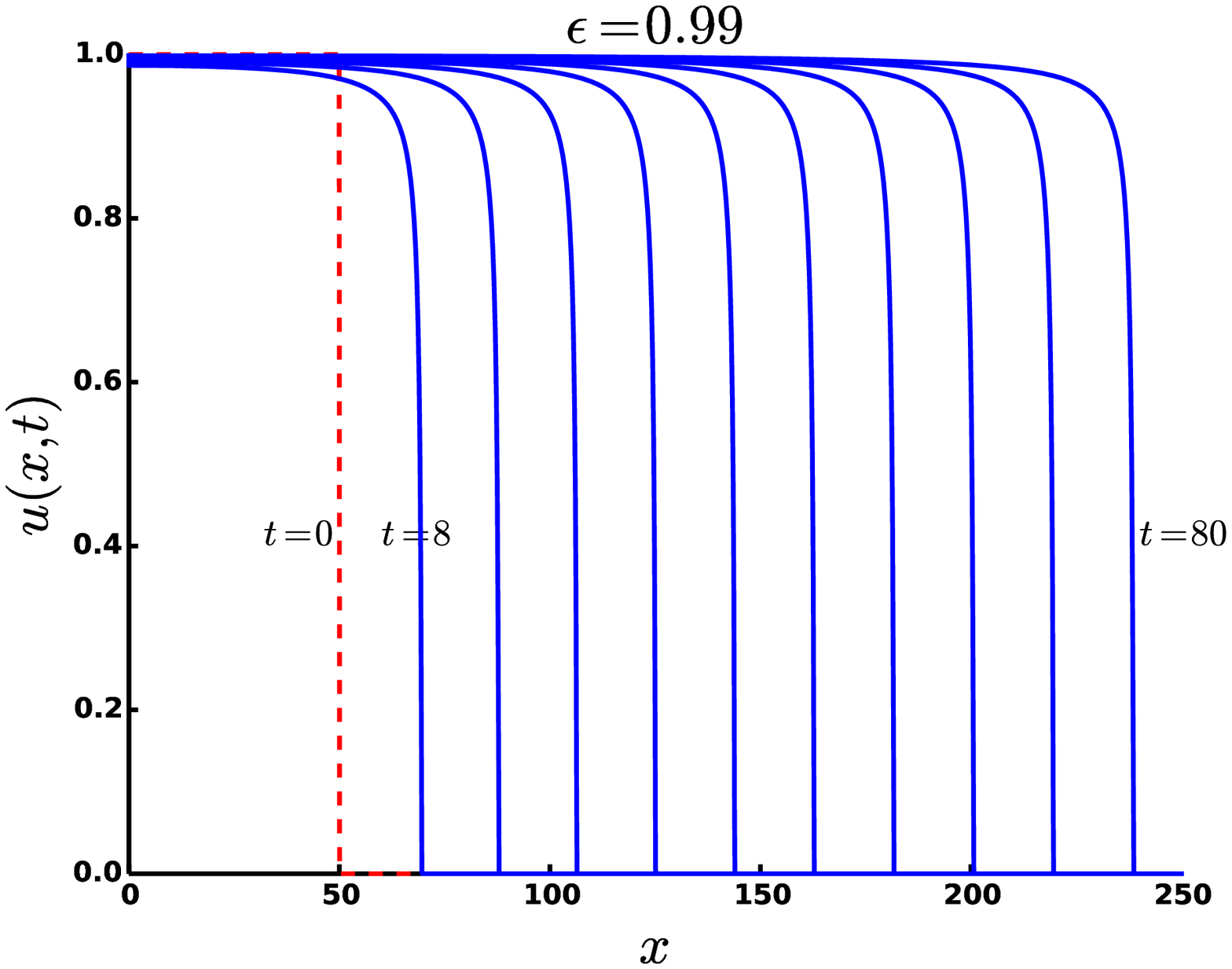}}
\end{subfigure}

\caption{\label{fig:Density}
(Color online) The demonstration of density profiles, $u(x,t)$, evolving from $t=0$ to $t=80$, obtained by using the numerical method. The dashed lines represent the initial density profiles. The data are shown for every $t=8$.
}
\end{figure}

\begin{figure}[ht]
\begin{subfigure}{0.5\textwidth}
\centerline{\includegraphics[width=\columnwidth]{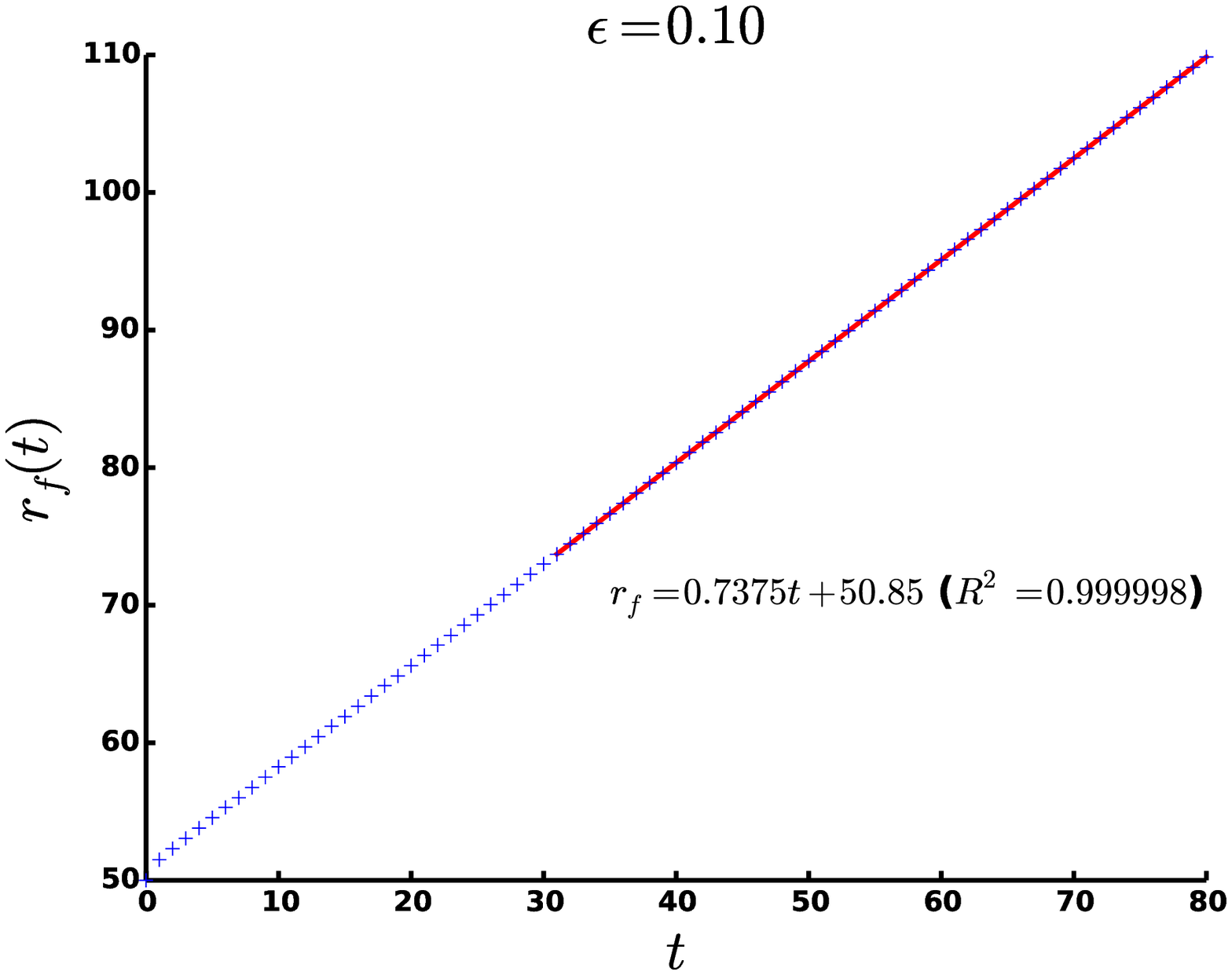}}
\end{subfigure}
\begin{subfigure}{0.5\textwidth}
\centerline{\includegraphics[width=\columnwidth]{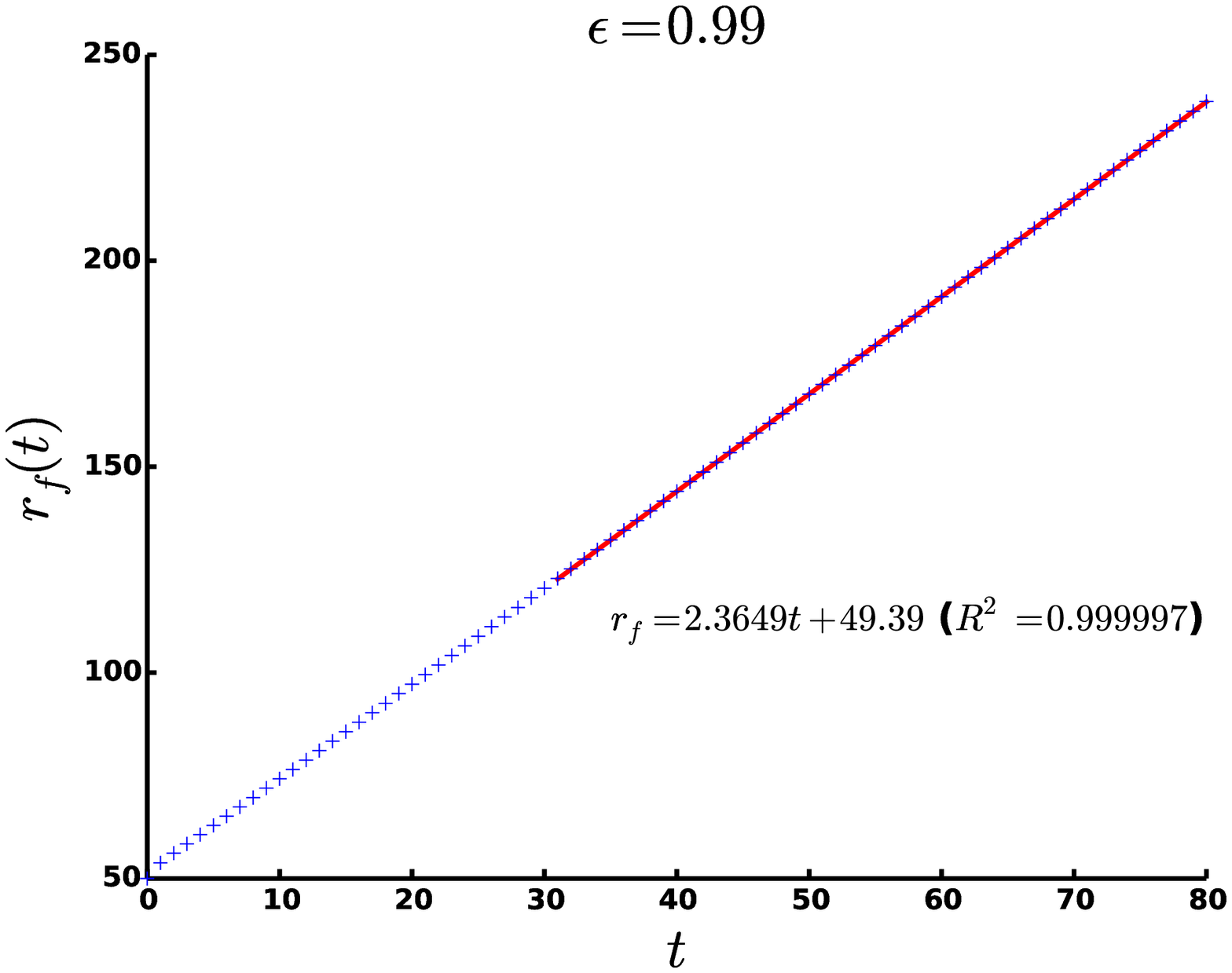}}
\end{subfigure}

\caption{\label{fig:Front}
(Color online) The front position versus time, corresponding to the numerical density profiles in \Fig{fig:Density}, from $t=0$ to $t=80$. The data are shown for every $t=1$. The markers represent numerical values and the solid lines represent the fitting lines for the last 50 data points. $R^2$ is the correlation coefficient.
}
\end{figure}

\begin{figure}[ht]
\centerline{\includegraphics[width=\columnwidth]{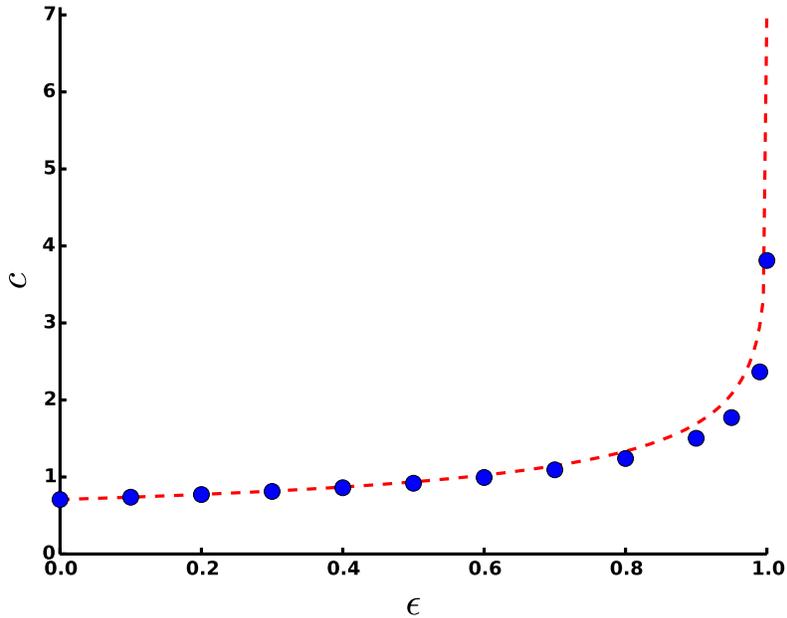}}
\caption{\label{fig:Speed}
(Color online) The front speed versus the packing fraction, $\epsilon$. The dashed line represents the analytical curve generated from \eq{eq:wave_vel_approx} and the circle markers represent the numerical results.
}
\end{figure}

The demonstration of the density profiles, obtained from the numerical method, is shown in \Fig{fig:Density} for dilute systems ($\epsilon=0.10$) and dense systems ($\epsilon=0.99$). It was observed that the density profile evolved with the sharp traveling wave with unchanged shape. The front position, $r_f(t)$, was determined by the first position where the density fell to zero. Due to numerical deviation, we measured the first position where the density was ≤ $1\times 10^{-6}$, or $u(r_f,t) \leq 1\times 10^{-6}$. The front positions were collected for every $t=1$. To avoid the transient effects of the initial stage, the last 50 data points were selected for fitting with the linear equation, $r_f = ct + r_0$. The corresponding front positions of the density profiles in \Fig{fig:Density}, as a function of time, were fitted well using the linear equation, as demonstrated in \Fig{fig:Front}. This implied that the density propagated with constant front speed, which was equal to the slope of the linear equation. We checked the accuracy of our algorithm by considering the front speed under conditions of $\epsilon=0$. In this case, the numerical front speed was equal to 0.7074, which displayed an error of ~0.04\% of the exact value ($c_0=1/\sqrt{2}\approx 0.7071$ \cite{Newman1980, Newman1983, Murray1989, Sanchezgarduno1994}). Finally, we explored the dynamics of bacterial populations in a closely packed regime. We set $\epsilon=0.999999$, in order to avoid dividing by zero for the factor $1/(1-\epsilon u)^2$ when $u=1$. In a closely packed system, the numerical front speed was equal to 3.8115, which was less than the analytically predicted value due to the inaccuracy of the approximate solution. The plot of the numerical front speed versus the packing fraction, as compared with the analytical curve generated from \eq{eq:wave_vel_approx}, is shown in \Fig{fig:Speed}. We found that the front speed increased with the packing fraction and reached a finite value as $\epsilon\to 1$. The analytical results agreed with the numerical data for the small packing fraction ($\epsilon \ll 1$), since the correction of our analytical solution was only $\Order{\epsilon^2}$. 

Finally, we compared our theoretical results to experimental evidence. From the experiments \cite{Sokolov2007, Rabani2013}, the average (or typical) velocity dependence upon the packing fraction of bacterial suspensions was determined. Below a critical packing fraction $< 1$, the average velocity of bacteria increased with the packing fraction and reached the maximum value at the critical packing fraction \cite{Sokolov2007, Rabani2013}. Above this critical point, the average velocity decayed to zero as the packing fraction approached one, due to the lack of free space. The increased front speed relative to the packing fraction observed in our model qualitatively agrees with the experimental observations under the former conditions. Their observations under the latter conditions were not observed in our results, given that the front speed in our model reached the maximum value when the packing fraction equalled one, which represents the closest packing fraction for a one-dimensional hard-rod system. Nevertheless, our data showed that the numerical front speed in a closely packed regime increased by a factor of $\sim 5$ relative to the dilute regime, which qualitatively agrees with experimental observations \cite{Sokolov2007, Rabani2013} showing increases in average velocity by a factor of $\sim 3$ in suspensions of spherical-shaped bacteria \cite{Rabani2013} and in typical velocity by a factor of $\sim 5$ in suspensions of rod-shaped bacteria \cite{Sokolov2007}.

\section{Conclusion \label{sec:Conclusions}}
This study demonstrated the effect of mechanical interactions between cells based on the spreading of bacterial populations by employing a continuum-mechanics modeling approach. In dense colonies, bacterial migration is dominated by hard-core repulsion between cells, which causes exclusion processes. The analytical and numerical results revealed that the expansion speed of bacterial colonies was enhanced by the exclusion effect and dependent upon the cell-packing fraction. These findings are qualitatively consistent with experimental evidence. 

\section*{Acknowledgment}
This research was supported by the TRF Grant for New Researchers (Grant No. TRG5780037), funded by The Thailand Research Fund and University of Phayao. 

\appendix
\section{Nonstandard fully implicit finite-difference scheme \label{sec:numer}}
We define the discrete density as $u^{n}_j = u(x_j,t_n)$, where $x_j=j\delta x$, $t_n=n\delta t$, $\delta x$ is a spacing step, $\delta t$ is a time step, $j \in \lbrace0,1,2,\ldots,J\rbrace$, $n \in \lbrace0,1,2,\ldots,N\rbrace$, and $J$ and $N$ are integers. We then rewrite \eq{eq:cell_RD_dimensionless_gen} as
\begin{equation}\label{eq:RD_numer1}
\frac{\partial u^{n+1}_j}{\partial t} \approx \frac{\partial}{\partial x} \left(M^n_j\frac{\partial u^{n+1}_j}{\partial x} \right) + f^n_j u^{n+1}_j,
\end{equation}
where $M^n_j = M(u^{n}_j) = u^{n}_j/\left(1-\epsilon u^{n}_j\right)^2$ and $f^n_j = 1-u^{n}_j$. Using the standard discretized scheme for the differential operators, we obtain
\begin{eqnarray}\label{eq:RD_numer2}
\lefteqn{
\frac{u^{n+1}_j - u^{n}_j}{\delta t} = \frac{1}{\delta x} \left(M^{n}_{j+1/2}\frac{\partial }{\partial x} u^{n+1}_{j+1/2} \right. } \nonumber \\ &&
\left. - M^{n}_{j-1/2}\frac{\partial }{\partial x} u^{n+1}_{j-1/2}\right) 
+ f^n_j u^{n+1}_j .
\end{eqnarray}
We discretize the remain gradient terms in \eq{eq:RD_numer2} and then have
\begin{eqnarray}\label{eq:RD_numer3}
\lefteqn{
\frac{u^{n+1}_j - u^{n}_j}{\delta t} = \frac{1}{\left(\delta x\right)^2} \left[M^{n}_{j+1/2}\left(u^{n+1}_{j+1}-u^{n+1}_{j}\right) \right. 
} \nonumber \\ &&
\left. - M^{n}_{j-1/2}\left(u^{n+1}_{j}-u^{n+1}_{j-1}\right)\right] 
+ f^n_j u^{n+1}_j.
\end{eqnarray}
The migration coefficient at the mid-grid can be computed by
\begin{eqnarray}
M^{n}_{j-1/2} &=& \frac{1}{2}\left(M^{n}_{j-1}+M^{n}_{j}\right), \\
M^{n}_{j+1/2} &=& \frac{1}{2}\left(M^{n}_{j}+M^{n}_{j+1}\right).
\end{eqnarray}
Noting that the correction of \eq{eq:RD_numer3} is $\Order{\delta{t}, \left(\delta{x}\right)^2}$. After rearranging \eq{eq:RD_numer3}, we have
\begin{equation}\label{eq:matrix1}
\alpha^n_j u^{n+1}_{j-1} + \theta^n_j u^{n+1}_j + \beta^n_j u^{n+1}_{j+1} = u^n_j,
\end{equation}
where 
\begin{eqnarray}
\alpha^n_j &=& -\mu M^{n}_{j-1/2}, \nonumber\\
\beta^n_i &=& -\mu M^{n}_{j+1/2}, \nonumber\\
\theta^n_j &=& 1  - \delta t f^{n}_j + \mu\left(M^{n}_{j-1/2} + M^{n}_{j+1/2}\right), \nonumber\\
\mu &=& \delta t/\left(\delta x\right)^2 .
\end{eqnarray}
We impose the zero-flux condition at the boundary grid, saying $\Omega$, that  $\left. \frac{\partial u}{\partial x} \right|_\Omega = 0$ or $\frac{u^n_{\Omega+1}-u^n_{\Omega-1}}{2\delta x} = 0$. Consequently, $u^n_{\Omega-1} = u^n_{\Omega+1}$ and $M^n_{\Omega-1/2} = M^n_{\Omega+1/2}$. We then rewrite \eq{eq:matrix1}, subjected to the zero-flux boundary condition, in matrix form:
\begin{equation}\label{eq:matrix_eq}
\textbf{A}^n \cdot \textbf{U}^{n+1} = \textbf{U}^{n},
\end{equation}
where
\begin{equation}\label{eq:matrix_M}
\textbf{A}^n = \left[
\begin{array}{ccccc}
\theta^n_0 & 2\beta^n_0      & \cdots             &  \cdots        & 0 \\
\alpha^n_1 & \theta^n_1      & \beta^n_1          &            	   & \vdots \\
\vdots     & \ddots           & \ddots             &  \ddots        &  \vdots    \\
\vdots     &                  & \alpha^n_{J-1}   & \theta^n_{J-1}  & \beta^n_{J-1} \\
0          & \cdots	          &      \cdots       & 2\alpha^n_{J}  & \theta^n_{J}                                 \end{array} 
\right],
\end{equation}
and
\begin{equation}\label{eq:matrix_U}
\textbf{U}^{n} = \left[
\begin{array}{cccccc}
u^{n}_0 & u^{n}_1 & u^{n}_2 & \cdots & u^{n}_{J}
\end{array} 
\right]^{\textnormal{T}} .
\end{equation}
According to the boundary condition, $\theta^n_0 = 1  - \delta t f^n_0 + 2\mu M^n_{1/2}$ and $\theta^n_{J} = 1 - \delta t f^n_{J} + 2\mu M^n_{J-1/2}$. The numerical density is obtained by solving the matrix equation (\eq{eq:matrix_eq}) iteratively.

To find the stability condition of this numerical scheme, we use von Neumann stability analysis:
\begin{equation}\label{eq:stable1}
u^n_j = \left(\lambda\right)^n e^{ikj\delta x},
\end{equation}
where $\lambda$ represents the amplification factor and $k$ is the wave number \cite{NumericalRecipes}. Substituting \eq{eq:stable1} into \eq{eq:RD_numer3}, we obtain $\lambda^{-1} = 1 -\delta t f^n_j - \mu M^n_{j+1/2}\left(e^{ik\delta x}-1\right) + \mu M^n_{j-1/2}\left(1-e^{-ik\delta x}\right)$, which can be approximated further: 
\begin{equation}\label{eq:stable2}
\lambda \approx \left[ 1 -\delta t f^n_j + 4\mu M^n_j \sin^2\left(k\delta x/2\right) + \Order{\delta x}\right]^{-1}.
\end{equation}
A stable and temporal non-oscillated numerical solution requires that $0<\lambda \leq 1$ \cite{Eberl2007}. According to $0\leq f^n_j \leq 1$ and $0\leq M^n_j < \infty$, without the growth term, ($f^n_j$), this algorithm is unconditionally stable as long as $\delta x \ll 1$ \cite{NumericalRecipes}. With the growth term, the solution slowly grows to a finite value as long as $\delta t \ll 1$. Based on \eq{eq:stable2}, this algorithm is adequately stable for this type of problem.


\bibliography{DDRDE_ref}

\end{document}